\documentclass[aps,prb,twocolumn,floatfix,superscriptaddress]{revtex4-1}
\usepackage[version=3]{mhchem} 
\usepackage[T1]{fontenc}       
\usepackage{graphicx}
\usepackage{float}
\usepackage{xspace}
\usepackage{amsmath}
\usepackage{comment}
\usepackage{mhchem}
\usepackage{tabularx}

\usepackage{verbatim}
\usepackage[table,xcdraw]{xcolor}

\usepackage{times}
\usepackage{ulem}

\usepackage{lineno,hyperref}
\modulolinenumbers[5]

\newcommand{\be}{\begin{equation}}
\newcommand{\ee}{\end{equation}}
\newcommand{\ba}{\begin{eqnarray}}
\newcommand{\ea}{\end{eqnarray}}

\def \cm {cm$^{-1}$\xspace}
\def \ctw {$^{12}$C\xspace}
\def \cth {$^{13}$C\xspace}

\begin{document}


\title{Real Space Raman Spectroscopy of Graphene Isotope Superlattices}

\author{Eric Whiteway}
\address{Department of Physics, McGill University, Montr\'eal, Canada H3A 2T8}
\author{Martin Lee}
\address{Department of Physics, McGill University, Montr\'eal, Canada H3A 2T8}
\address{Kavli Institute of Nanoscience, Delft University of Technology, Lorentzweg 1, 2628 CJ, Delft,
	The Netherlands}
\author{Michael Hilke}
\address{Department of Physics, McGill University, Montr\'eal, Canada H3A 2T8}
\address{Center for the Physics of Materials (CPM and RQMP)}

\begin{abstract}

We report the Raman spectroscopy of  \ctw/\cth graphene isotope superlattices synthesized by chemical vapour deposition. At large periods the Raman spectra corresponds to the sum of the bulk \ctw and \cth contributions. However, at small periods we observe the formation of mixed \ctw/\cth modes for Raman processes that involve two phonons, which results in the tripling of the 2D and 2D$'$ Raman peaks. This tripling can be well understood in the framework of real space Raman spectroscopy, where the two emitted phonons stem from different regions of the superlattice. The intensity of the mixed peak increases as the superlattice half period approaches the mean free path of the photo-excited electron-hole pairs. By varying the superlattice period between 6 and 225 nm we have a direct measure of the photo-excited electron mean free path, which was found to be 18 nm for suspended graphene and 7 nm for graphene on SiO$_2$ substrates.
\end{abstract}

\maketitle

\section{Introduction}

Raman spectroscopy is a powerful technique to measure vibrational energies through inelastic photo-excited electron scattering processes via the emission or absorption of phonons. These processes are typically viewed in momentum space, where momentum conservation plays an important role in the electron-phonon scattering processes. However, in the presence of
short ranged spatial variations of the phonon modes, real space considerations become important. This is particularly relevant, when the electronic degrees of freedoms are spatially invariant as opposed to the vibrational properties. For instance, this is true in crystals, where the isotopes of the atoms have a spatial dependence, since the different masses will modify the vibrational properties, but not the electronic ones. Particularly interesting, is the case where in a single Raman process, it is possible to generate two phonons from two regions with different atomic masses. This would lead to additional new second order Raman lines, which we discuss below.

The Raman spectrum of graphene is quite unique due to the prominence of two-phonon Raman processes. Indeed, the strongest Raman peak in pure graphene is the two phonon 2D Raman peak \cite{thomsen2000double,maul04,ferr06}. Isotope superlattices composed of alternating bands of \ctw and \cth result in a spatial variation of the phonon local density of states (LDOS) while preserving the crystal structure and electronic properties of graphene. They are therefore an ideal platform to investigate real space Raman processes. In this work we report the synthesis of isotopic graphene superlattices with periods as low as 6 nm and the resulting structure dependent tripling of the two phonon Raman peaks. We show that this peak structure is caused by a non-local Raman processes involving the emission of two spatially separated phonons and provides a direct measure of the mean free path of photo-excited electrons. Thus, Raman scattering provides a unique tool to probe the spatial variation of phonon modes at scales much smaller than the optical wave length.   

Graphene isotopic superlattices have been extensively studied theoretically using molecular dynamics \cite{ouya09,mu15,feli18,xie17,gu18,davi17} with a focus on thermal conductivity and acoustic phonons as opposed to the effect of the superlattice structure on the optical phonon modes. However there have not been experiments on isotopic superlattices until very recently \cite{whiteway2020thermal}, where a strong strong suppression of the thermal conductivity was observed due to the isotope hetero interface.

Interest in superlattices is primarily motivated by the expected reduction in thermal conductivity\cite{yao87,yash98,sim00}  and the unique properties of graphene\cite{novo04,bolo08,drag07} may make a graphene SL an ideal material for thermoelectric devices. The synthesis of nm scale graphene superlattices with tuneable interface density therefore represents an important advancement and we present a framework to directly characterize the isotope concentration and superlattice period in graphene and other 2D materials through Raman spectroscopy.

To understand the effects of spatial variations in the phonon modes it is important to consider the real space Raman picture (see figures \ref{diagram} and \ref{rsr_fig}), which we describe next.

\section{Real space resonant Raman processes}
\label{sec:rsr}
In general, Raman spectroscopy of graphene will identify a number of phonon energies at well defined regions in momentum space. For instance, the so-called G-peak corresponds to an emission (Stokes) or absorption (anti-Stokes) of phonons at the $\Gamma$ point in the Brillouin zone. The most prominent peak (2D) corresponds to two phonons close to the K (or K') points along the in-plane Transverse Optical (iTO) phonon mode with wavenumber determined by the laser energy \cite{mala09}.

\begin{figure}[h!]
\includegraphics[width=0.6\columnwidth]{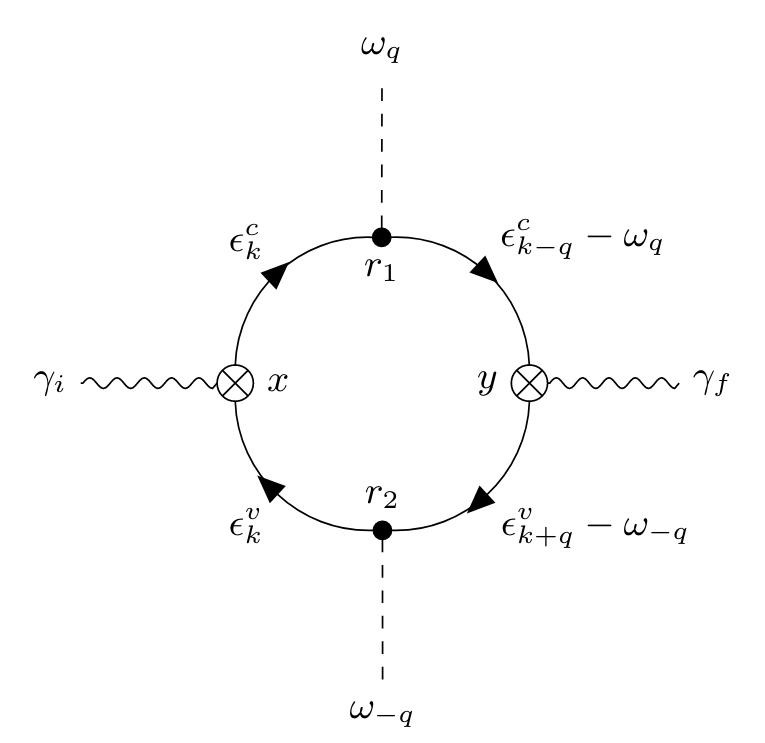}
  \caption{A two phonon Stokes Feynman diagram for the 2D or 2D$'$ resonant Raman scattering process. $\gamma_{i,f}$ are the incoming and Raman shifted outcoming photon energies, $\omega_{\pm q}$ are the emitted phonon energies, and $\epsilon_k^{c,v}$ are the electron and hole energies.}
  \label{diagram}
\end{figure}

\begin{figure*}[htbp]
\includegraphics[width=\textwidth]{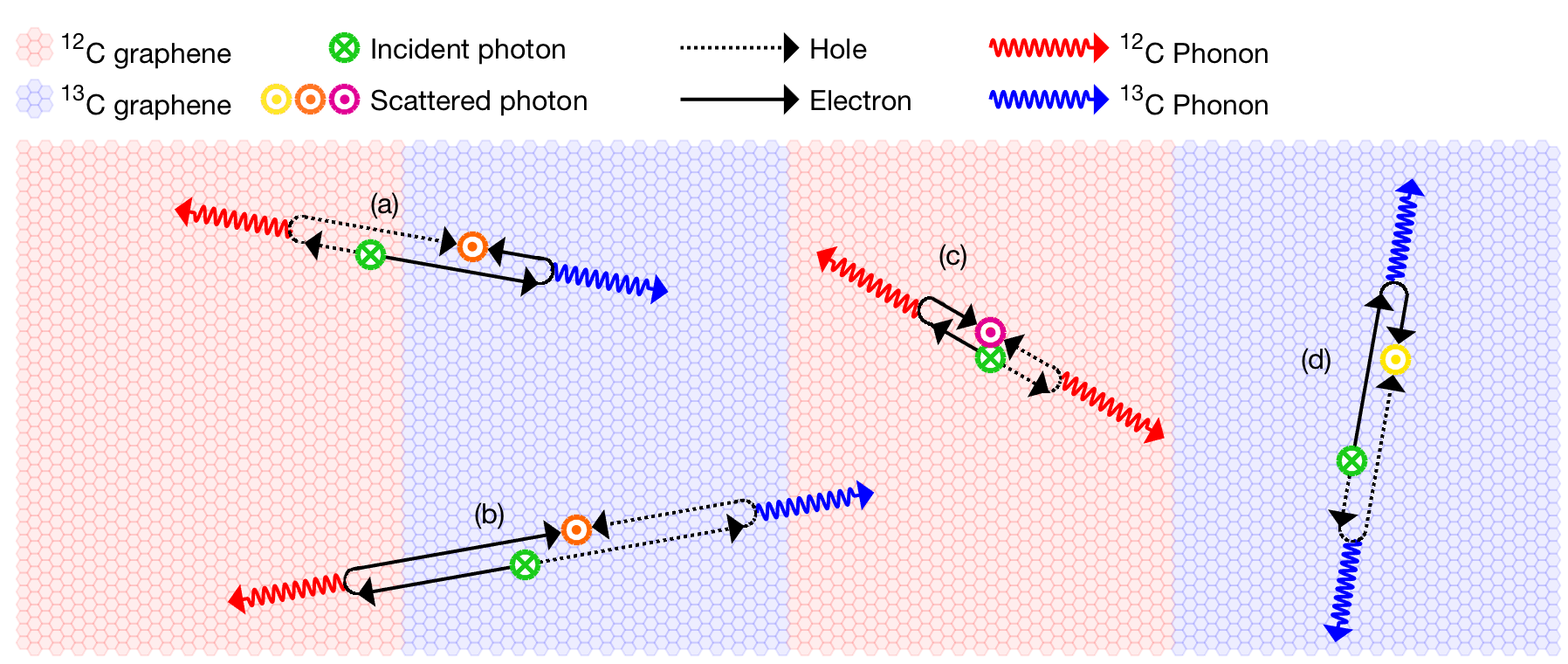}

\caption{Diagram of possible real space Raman two phonon processes in a \ctw/\cth superlattice with energies given by a) $\omega^{13,12}$, b) $\omega^{12,13}$, c) $\omega^{12,12}$ and d) $\omega^{13,13}$. Solid and dotted lines represent the semiclassical trajectory of the photo-excited electron and hole respectively. Note that, as described in section \ref{sec:rsr},the energy of the scattered photon is uniquely determined by the position of phonon emission $r_1$ and $r_2$ and not the position of the incident of emitted photon $x$ and $y$.}
\label{rsr_fig}
\end{figure*}

In graphene, the strong 2D and narrow 2D$'$ peaks in Stokes Raman spectroscopy arise from the emission of two non-zero momentum phonons. Their higher amplitude than their single phonon counterpart (D and D') can be explained by their double or triple resonant structure \cite{thomsen2000double,vene11} or the sliding mechanism\cite{heller2016theory} due to the linear electronic dispersion. What is important in our context, is that the two phonons involved are spatially separated due to the finite momentum transfer between electron-holes and phonons. While many processes can contribute to the two phonon Raman scattering amplitude, the most important one is shown in figure \ref{diagram}. The Feynman diagram is drawn in the real space configuration, where a photo-excited electron is created at $x$ after a photon absorption of energy $\gamma_i$. The electron of energy $\epsilon_k^c\simeq\gamma_i/2$ is scattered by a phonon of energy $\omega_q$ and momentum $q$ at $r_1$, while the hole of energy $\epsilon_k^v=\epsilon_k^c-\gamma_i$ is scattered by a phonon of energy $\omega_{-q}$ and momentum $-q$ at $r_2$. The electron and hole recombine at $y$ to emit a photon of energy $\gamma_f=\gamma_i-\omega_q-\omega_{-q}$. 

This process and all the other relevant diagrams can be evaluated with standard diagrammatic techniques and summed up to obtain the full cross-section. However, we will restrict the discussion to the process shown in figure \ref{diagram}, which involves intermediate states that are all real electronic states (not virtual), which will give a dominant contribution\cite{vene11}. We can further treat the photo-excited electron hole pair quasi-classically, {\it i.e.}, as a localized wave packet with well defined energy and momentum. This approach \cite{zeyher1974calculation} was used extensively by Basko and co-workers in graphene \cite{bask08,ferr13} and in magneto-Raman \cite{faug10}. In this picture, the incoming light of energy $\gamma_i$ produces an electron and a hole at $x$ with opposite group velocity $v=\partial_k\epsilon_k$, where $\epsilon_k$ is given by the hallmark conical dispersion relation of graphene, $v\simeq 10^6 m/s$ and momentum $k=\frac{\gamma_i}{2v}$. The quasi-free electron and hole will follow their initial trajectories of opposite velocity until they scatter with other electrons or phonons. If the electron scatters with a phonon of momentum $q$ at $r_1$ and the hole scatters with a phonon of momentum $-q$ at $r_2$, they can eventually recombine at $y$ and emit a Raman shifted photon. This process is illustrated in figure \ref{rsr_fig}.
 
A necessary condition for recombination is that neither electron nor hole undergo another scattering event. However, this is generally quite likely, which leads to the well known suppression of Raman events. For the events that contribute to the Raman amplitude, if the electron is scattered by a \ctw phonon at $r_1$ and the hole is scattered by a \cth phonon at $r_2$ then the phonon emitted by the electron will have a different energy from the phonon emitted by the hole, yet both phonons will have opposite momenta. This would lead to a combination Raman 2D peak at an energy $\omega_{2D}=\omega^{12}_D+\omega_{D}^{13}$, where $\omega_{2D}$ is the measured Raman 2D peak shift and $\omega^{\alpha}_D$ the D-phonon energy for isotope $\alpha$. In general, there will be three possible energies for the 2D peak: $\omega_{2D}^{\alpha\beta}=\omega_D^\alpha+\omega_D^\beta$, where $\alpha$ and $\beta=12$ or $13$ as illustrated in figure \ref{rsr_fig} for an isotope superlattice. The typical separation between the electron and hole when they scatter with two phonons (not necessarily at the same time) is $d_{e-h}=|r_1-r_2|=\lambda$, where $\lambda$ is the electronic mean free path (MFP).

This real space picture allows us to conveniently estimate the relative strengths of each process by identifying the corresponding spatial location probabilities of the electron and hole. If $\tilde{r}_e(t)$ is the semiclassical trajectory of the electron, then the probability to emit a phonon at $r_1$ at time $t_1$ and to recombine at $y$ at time $t=t_1+t_2$ is proportional to $\sim e^{-t/\tau}$, where $\tau$ is the total scattering time. Equivalently, the probability for the hole with trajectory $\tilde{r}_h(t)$ to emit a phonon at $r_2$ at time $t_2$ and to recombine at $y$ at time $t$ with the electron is also $\sim e^{-t/\tau}$, where we assumed that the electron and hole have the same scattering time $\tau$. Therefore, the normalized joint probability for the electron to emit an $\alpha$ phonon and for the hole to emit a $\beta$ phonon is given by
\begin{widetext}
\be P^{\alpha\beta}(\tilde{r}_e, \tilde{r}_h)=\left(\frac{2}{\tau}\right)^2\int_0^\infty dt_1\int_0^\infty dt_2\,e^{-2(t_1+t_2)/\tau} M^\alpha[\tilde{r}_e(t_1)]M^\beta[\tilde{r}_h(t_2)],\ee
where $M^\alpha(r)=1$ if we have isotope $\alpha$ at position $r$ and zero otherwise. We have $M^{12}(r)+M^{13}(r)=1$. 

The relative integrated intensities of the 2 phonon Raman peaks is then given by summing over all the electron and hole trajectory pairs:
\be F^{\alpha\beta}=\frac{1}{N_{tot}}\sum_{\tilde{r}_e,\tilde{r}_h}P^{\alpha\beta}(\tilde{r}_e, \tilde{r}_h),\ee
where $F^{12,12}+F^{12,13}+F^{13,12}+F^{13,13}=1$.

In the simple case of a one dimensional isotope superlattice of period $L_s$ in the direction $x$ we have: 

\be F^{\alpha\beta}=\frac{2}{\pi\tau^2 L_s}\int_0^\infty dt_1\int_0^\infty dt_2\int_0^{L_s}dx\int_0^{2\pi}d\theta\,e^{-2(t_1+t_2)/\tau} M^{\alpha}[x+vt_1\cos(\theta)]M^\beta[x-vt_2\cos(\theta)].\label{maineq}\ee
\end{widetext}

The relative fraction of integrated intensities $F^{\alpha\beta}$ will depend on $L_s$ and the electronic mean free path $\lambda=v\tau$. In the case of the 2D and 2D$'$ modes the two  permutations (12,13) and (13,12) are degenerate in energy and we write the relative intensity of the Raman signal as: 
\be 
F^{\{12,13\}}=F^{12,13}+F^{13,12}
\ee
 
For $\lambda\gg L_s$ all $F^{\alpha\beta}$'s are equal, while for $\lambda\ll L_s$ we have $F^{\{12,13\}}\simeq \frac{4\lambda}{\pi L_s}$ and a crossover region when $\lambda\simeq L_s$. 

If we consider only the semiclassical trajectories of the electron-hole pair we find that the x-component of the separation, $x_{e-h}$ is described by the distribution:

\begin{equation}
P(x_{e-h})=\frac{8}{\pi \lambda^2} x K_1(\frac{2x}{\lambda})
\end{equation}

Where $K_n$ is the Bessel function of the second kind. In the case of a superlattice with interface density $I_d=1/L_s$ this leads to a dependence of $F^{12,13}$ given by:

\begin{equation}
F^{\{12,13\}}=\frac{1}{2}-\frac{4}{\pi^2}\sum_{n=1,3,5...}^{\infty} \frac{1}{n^2(1+n^2\pi^2\lambda^2 I_d^2)^{3/2}}
\end{equation}

The overall dependence can be well approximated numerically by:
\be
F^{\{12,13\}}\simeq\left[\left(\frac{\pi L_s}{4\lambda}\right)^{3}+f_0^{-3} \right]^{-1/3},
\label{Ialpha}
\ee

Where $f_0$ is a constant determined by the duty cycle of \ctw and \cth in the superlattice structure with $f_0=1/2$ in the case that the length of the \ctw and \cth regions are equal.

Equation \ref{Ialpha} can be used to extract $\lambda$ from $L_s$ as discussed in the following sections.

\section{Combination \ctw/\cth Raman peaks}

\ctw/\cth graphene isotope superlattices with periods ranging from 225 to 6 nm were prepared by CVD. The synthesis and characterization of these samples are described in sections \ref{sec:synth} and \ref{sec:char}.

\begin{figure}[htb]
\includegraphics[width=3.4in]{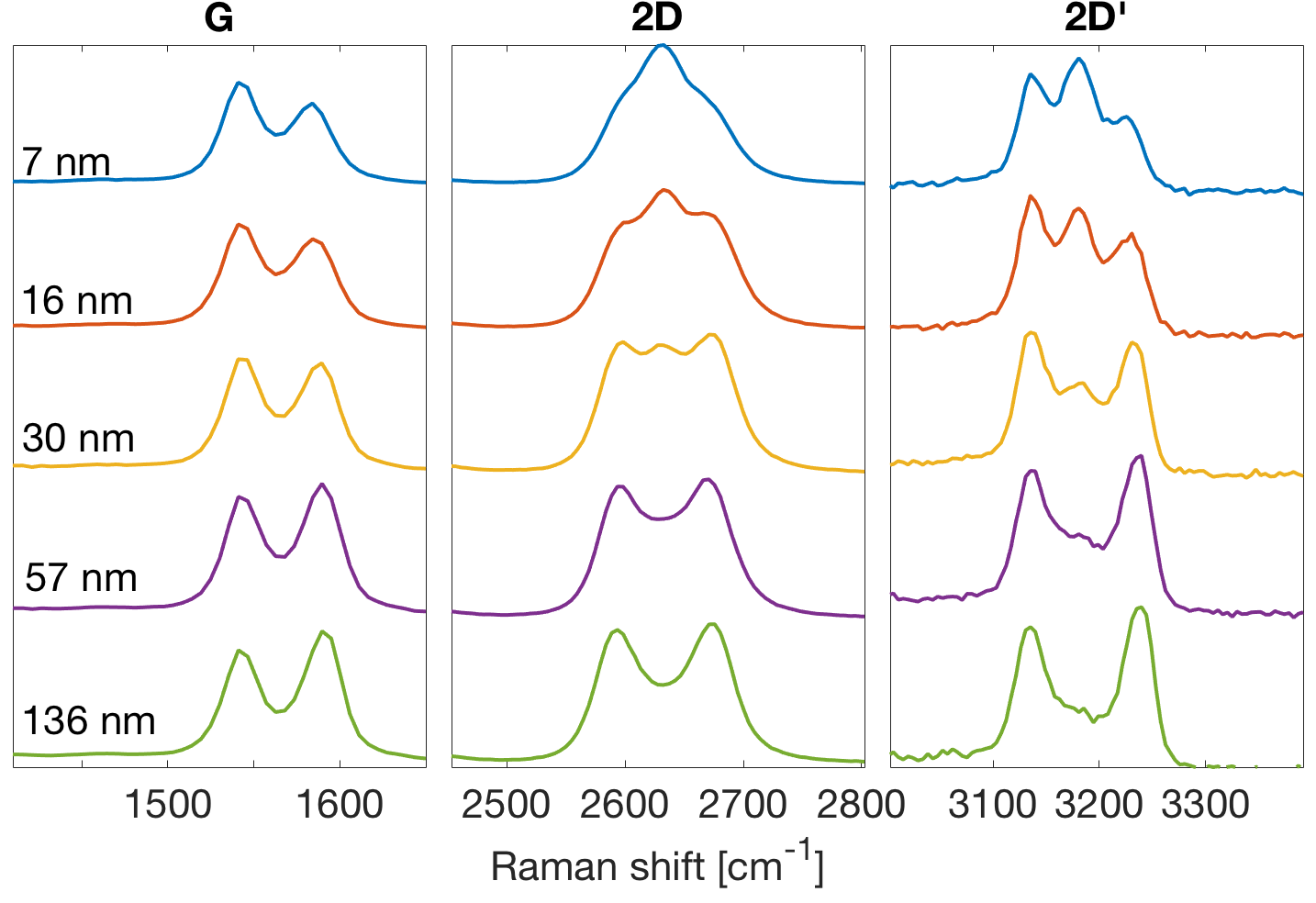}
\caption{Raman spectra as a function of superlattice period, $L_s$. Spectra are averaged over 10-20 spots taken from a single sample with a wide range of periods.}
\label{spectra}
\end{figure}

At superlattice periods greater than 100 nm we observe in figure \ref{spectra}, as expected, double peaks for each Raman mode in graphene corresponding to \ctw and \cth graphene bulk Raman spectrum. At small periods we also observe the formation of a third peak in the 2D and 2D$'$ modes as seen in figure \ref{spectra}. The additional middle Raman peaks arise from processes involving two spatially separated phonons, one \ctw and one \cth phonon as described in detail in section II. The frequencies and relative intensities of the peaks are extracted by fitting the 2D and 2D$'$ modes with a triple lorentzian peak structure. The Raman shift of this middle peak is the average of the \ctw and \cth Raman peaks $\omega_{2D}^{\{12,13\}} = \frac{1}{2}(\omega_{2D}^{12,12} +\omega_{2D}^{13,13})$ and the intensity $F^{\{12,13\}}$ scales with decreasing SL period as given by equation \ref{Ialpha}.

\begin{figure}[htb]
    \includegraphics[width=3.2in]{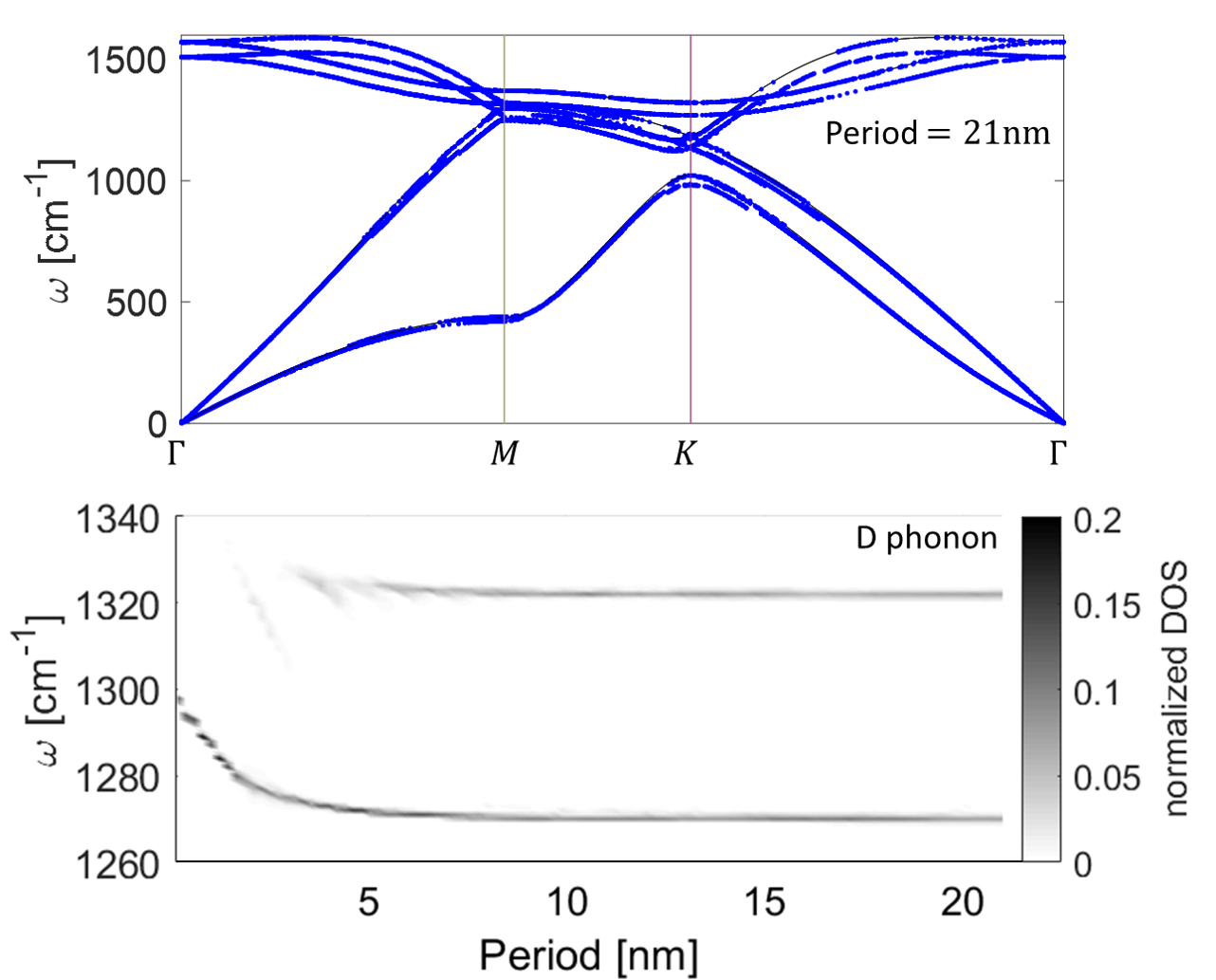}
	\caption{Calculated superlattice phonon dispersion shown for a \ctw/\cth superlattice of period 21 nm (blue dots). The solid lines show the 2D dispersion for pure \ctw. The bottom graph shows the DOS for the D peak as a function of superlattice period.\cite{hilke2020phonon}}
	\label{disp}
\end{figure}

In the case of the one phonon G process we don't observe the formation of a third peak. The observation that the third peak is only present for 2 phonon Raman processes as well as the lack of any features in the calculated phonon DOS for $L_s>6$ nm (see figure \ref{disp}) corresponding to this intermediate peak strongly suggest that it results from a 2-phonon process involving one \ctw and one \cth phonon. It also precludes the possibility that this peak is simply the result of the underlying isotope distribution, since a distribution peaked at an isotope concentration $\rho=0.5$ would be evident in the G peak structure. 

While the 2D and 2D$'$ peaks involve two phonons on the same phonon branch (close to the K point for 2D and close to $\Gamma$ for 2D$'$), other combination peaks such as D'+D$^3$ and D+D", involve two different phonon branches \cite{ber12}, In this case $\omega^{12,13}\neq \omega^{13,12}$, which would lead to two additional phonon peaks as shown in the supporting information.

\begin{figure}[htbp]
\includegraphics[width=3in]{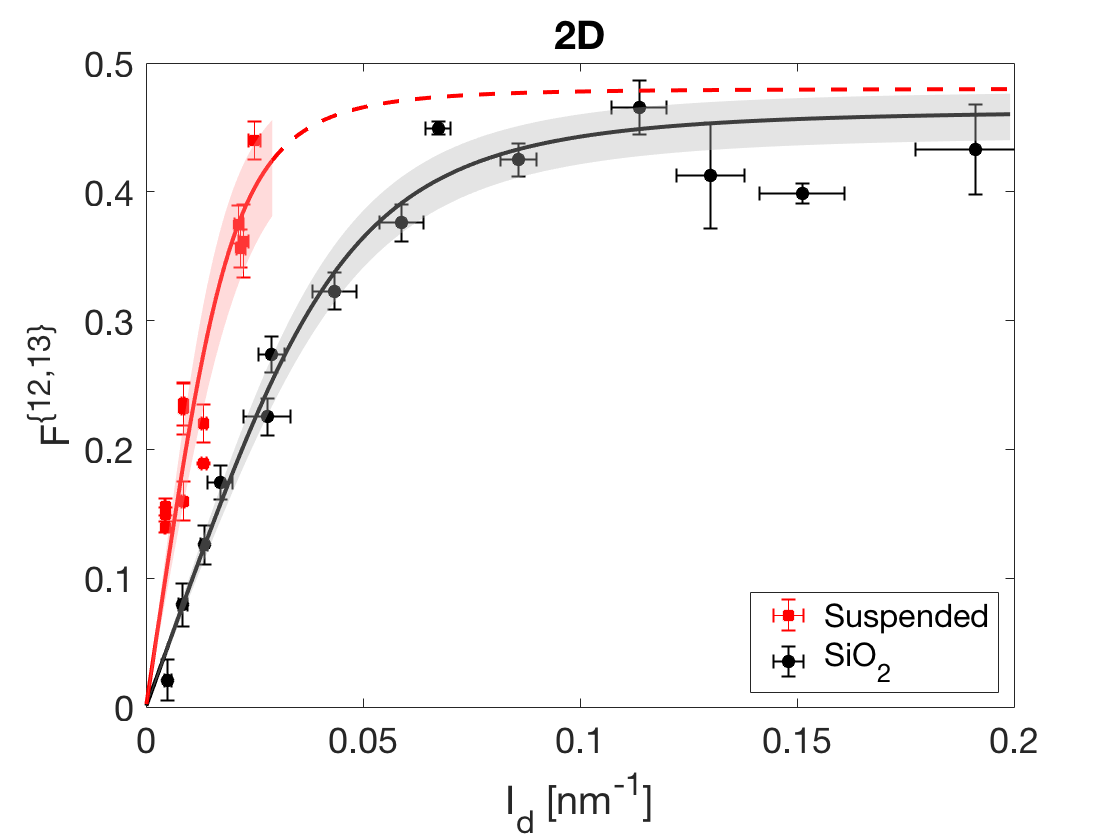}
\includegraphics[width=3in]{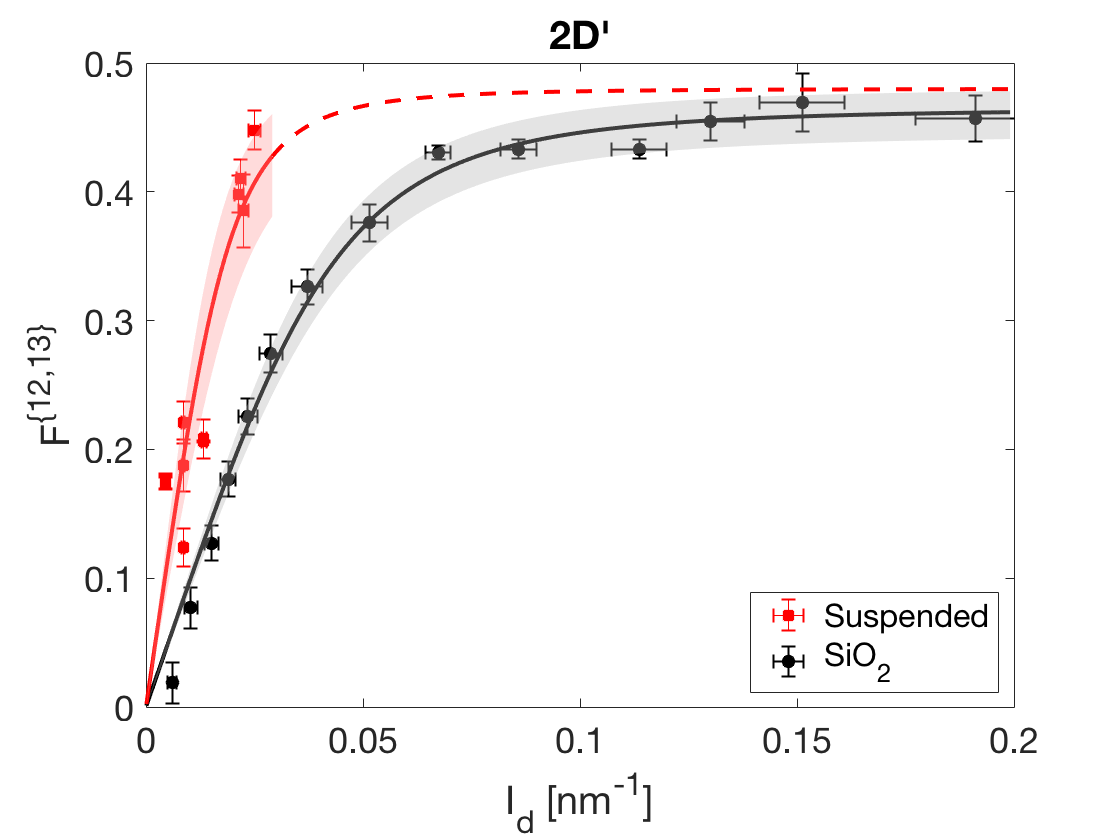}
\caption{Length dependence of the Relative intensity of mixed process Raman peak for the 2D and 2D$'$ mode. $F^{\{12,13\}}$ is plotted against the interface density $I_d=1/L_s$. We observe as expected at small values of $I_d$ a linear dependence with slope determined by $\lambda$. Error bars show the standard error. The data is fit to equation \eqref{Ialpha} and the shaded error shows the 95\% confidence interval of the fit parameters}
\label{hbz2dp}
\end{figure}

For the 2D and 2D$'$ Raman amplitudes, we find that the relative intensity $F^{\{12,13\}}$ of the mixed Raman peak increases with decreasing superlattice period approximately as $1/L_s$. This is shown in figure \ref{hbz2dp}. The dependence is well fitted by equation \eqref{Ialpha}, which depends on the ratio of the superlattice period $L_s$ and the MFP. 

\section{Photo excited electron mean free path}

The incoming Raman laser beam excites the electrons by the energy of the photon. With momentum conservation in the Dirac cone electron dispersion, most photo-excited electrons will have an energy close to $\epsilon_L/2$ from the K point, where $\epsilon_L$ is the incoming photon energy. These photo excited electrons will rapidly decay to lower energies by inelastic scattering with other electrons and phonons. Time resolved experiments in graphene show that this decay starts to happen in the 10 fs range \cite{breusing2011ultrafast}. Experiments and simulations seem to indicate that the initial electron-electron scattering is followed by electron-optical phonon scattering spanning 10-300 fs \cite{breusing2011ultrafast,tomadin2013nonequilibrium,brida2013ultrafast}. Time resolved Raman spectroscopy experiments in graphite have shown the full building of the G phonon Raman mode to be below 300fs with an initial build-up within 20 fs \cite{yan2009time,ishioka2008ultrafast}.

For the photo-excited electrons and holes, the total inelastic mean free time is given by $\tau_{tot}^{-1}=\tau_{e-ph}^{-1}+\tau_{e-e}^{-1}$, where  $\tau_{e-e}^{-1}$ is the electron-electron scattering rate while $\tau_{e-ph}^{-1}$ is the electron-phonon scattering rate. For hot electrons and photo-excited electrons in graphene $\tau_{tot}$ was calculated to be in the 10-120fs range and dependent on the Fermi energy \cite{tse2008ballistic,song2013photoexcited}. The time resolved experiments discussed above, are consistent with a shorter $\tau_{e-e}$ compared to $\tau_{e-ph}$. In this case only a small fraction of the Raman photo excited electrons will generate a Raman phonon.

The Raman analysis in the real space picture, gives us a direct measurement of the MFP of photo-excited electrons, which is connected to the total scattering time by $\lambda=v\tau_{tot}$. Using equation \eqref{Ialpha} we can fit the relative intensities as a function of superlattice period to obtain $\lambda$. This is shown in figure \ref{hbz2dp} for both suspended graphene and graphene supported on SiO$_2$ and is in good agreement with the experimental data. A least squares fit  gives  $\lambda$ of 18 nm $\pm$ 4 nm in suspended graphene compared to 7.4 nm $\pm$ 0.6 nm in graphene on SiO$_2$.

While there are no other direct experimentally measured mean free paths of the photo-excited electrons or holes in graphene various measurements have been made of the electronic linewidth and the exciton lifetime, by time or angle resolved photo-emission spectroscopy or Raman experiments and report values ranging from 28-100 meV. Table \ref{reftable} summarizes results measuring electronic broadening, excitation lifetime and mean free path of carriers in graphene and HOPG. The corresponding value of $\lambda$ is obtained from $\lambda=\hbar v/2\gamma$. Electronic broadening is reported depending on the reference as one of $\gamma$, $2\gamma$ or $4\gamma$ and here are standardized as $2\gamma$. Our measured values of $\lambda$ correspond to values of $2\gamma\simeq 36$ meV for suspended graphene and $\sim$89 meV for graphene on SiO$_2$ which are similar to the other results reported in the literature.

\begin{table}[htbp]
\label{reftable}
\begin{tabularx}{\columnwidth}{rrXr}
$2\gamma$ [meV]& $\lambda$ [nm] &  Technique & Reference \\ \hline
    \rowcolor[HTML]{EFEFEF}[\tabcolsep] 
\bf{100}   & 6.6 &  ARPES & \cite{bost07}  \\
89   & \bf{7.4} &  Raman-SL (supp.)& this work  \\
    \rowcolor[HTML]{EFEFEF}[\tabcolsep] 
\bf{66}   & 10 &  Raman & \cite{bask09}  \\
\bf{54}    & 12  & Magneto-Raman  & \cite{faug10}  \\
    \rowcolor[HTML]{EFEFEF}[\tabcolsep] 
\bf{48}    &  14 & Raman  & \cite{vene11}\\
36   & \bf{18} &  Raman-SL (susp.)& this work  \\
    \rowcolor[HTML]{EFEFEF}[\tabcolsep] 
   \bf{ < 33}   & > 20 & ARPES (epitaxial) & \cite{spri09}  \\
\bf{28}    &  24 & TRPES (HOPG)  & \cite{moos01}\\


\end{tabularx}
\caption{Comparison with previous literature values of electronic linewidth in graphene. In order of increasing MFP. }
\end{table}

The total probability of resonant two phonon processes is proportional\cite{bask08} to $\lambda^2$. Therefore if we consider the Raman non-resonant G peak intensity to be independent of scattering rate then we expect the ratio $I_{2D}/I_G \propto \lambda^2$. The ratio $I_{2D}/I_G$ has previously been shown increase for suspended graphene vs. graphene on SiO$_2$ which was attributed to a decrease in charged impurities \cite{ni09}. Similarly we find that suspending graphene increases both the measured value of $\lambda$ and $I_{2D}/I_G$ compared to the results on SiO$_2$ substrates. For suspended graphene the ratio $I_{2D}/I_G\simeq 6$ whereas for supported graphene we measure $I_{2D}/I_G\simeq 4$. (see SI for details)

\section{Dependence on polarization}
\label{sec:polar}

\begin{figure}[htbp]
	\includegraphics[width=.99\columnwidth]{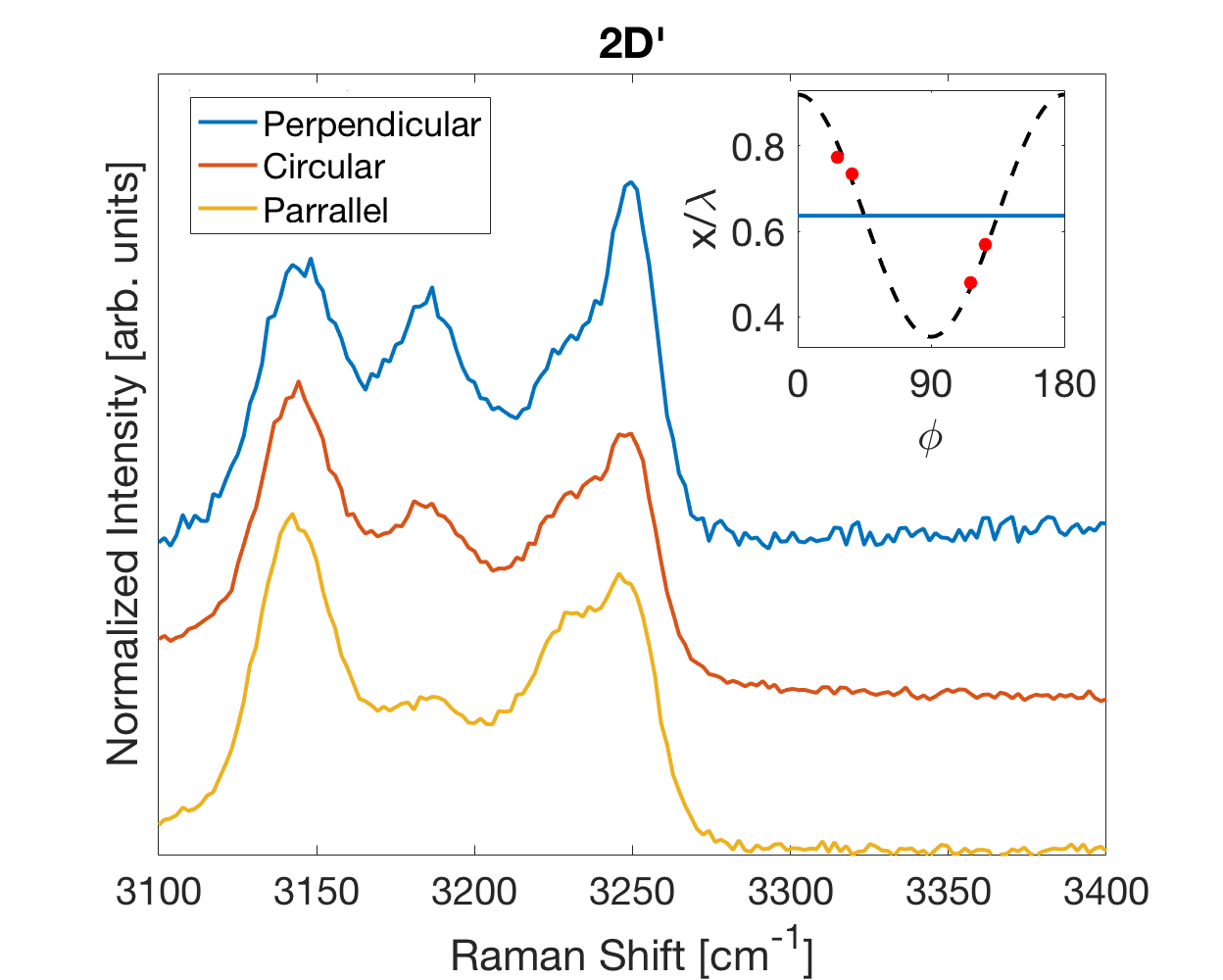}

	\caption{Polarization dependence: 2D$'$ spectrum for circular and linear poarizations approximately perpendicular and parallel to the mass periodicity. Inset: theoretical polarization dependence of $x/\lambda$ given by equation \ref{poldep}, measured polarizations are shown in red.}
	\label{polar}
\end{figure}

The real space Raman process described in section II will invariably lead to a dependence on the polarization of the incoming light with respect to the superlattice orientation, since the photo-excited electron-hole pair will more likely have a momentum perpendicular to the polarization. For an angle $\phi$ measured between the electric field polarization and the electron-hole pair momenta the probability of detecting a photon\cite{bask08}, corresponding to the backscattered electron-hole pair, varies as $(\sin\phi)^4$. Hence, electron-hole pairs with momenta in the direction of periodicity will more likely result in the emission of 12-13 phonon pairs and as a result the value of $F^{\{12,13\}}$ will vary as a function of polarization angle as shown in figure \ref{polar}. 

The magnitude of $F^{\{12,13\}}$ is varied by the parameter $x$, the component of the mean free path in the direction of periodicity, which we take to be $\frac{2}{\pi}\lambda$ for circularly polarized light. We can quantify the polarization dependence by considering the value of $x(\phi)$ as a function of polarization angle $\phi$ as:

\begin{equation}
\frac{x(\phi)}{\lambda}=\frac{8}{3\pi}\int_0^\pi \sin(\theta)[\sin(\theta-\phi)]^4d\theta
\label{poldep}
\end{equation}

The value of $F^{\{12,13\}}$ was measured for different linear polarizations and the corresponding value of $x$ was extracted by fitting to equation \eqref{Ialpha}. We take $\phi=0$ to be polarized perpendicular to the periodicity of the superlattice. This is shown in the inset of figure \ref{polar}.

\section{Synthesis of isotope superlattice}
\label{sec:synth}

Graphene was grown by low pressure chemical vapour deposition on commercially available 25 $\mu$m thick copper foils. During the growth phase \ctw-methane and \cth-methane were pulsed in an alternating sequence. The methane sources were respectively 99.99\% pure $^{12}$C methane or 99\% pure  $^{13}$C-methane (Sigma-Aldrich 490229).  The duration of the pulses was on the order of 1 second followed by a 2-4 second period with no methane flow in order to maintain high isotope concentration throughout the growth.

\begin{figure}[htbp]
\includegraphics[width=.9\columnwidth]{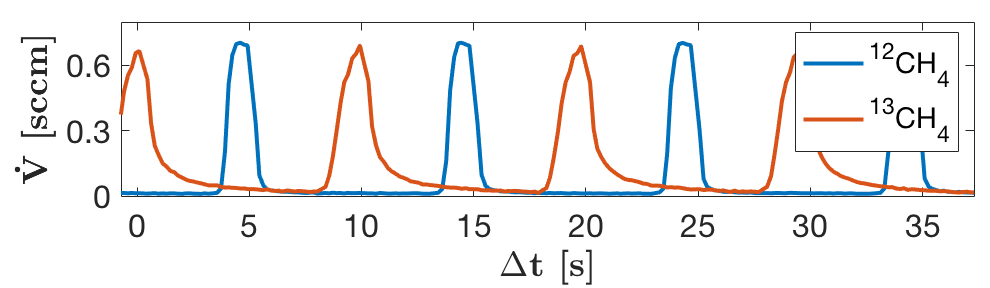}
\includegraphics[width=.9\columnwidth]{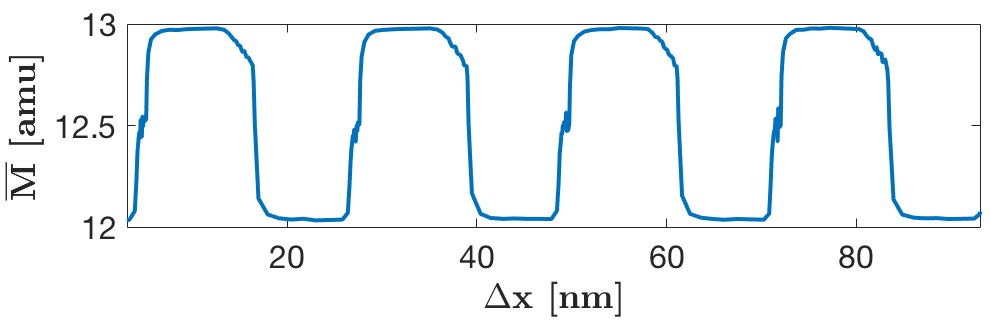}
\caption{growth log showing flow rates $\dot{V}$ of \ctw and \cth methane along with the corresponding atomic mass vs. distance where distance is calculated as $\dot{V}\Delta t$ and scaled to correspond to the measured growth rate and the average atomic mass is obtained from the isotopic methane concentration.}
\label{gasflows}
\end{figure}

Periodically a long (1 min) pulse of pure \ctw or \cth  methane was introduced which allows us to distinguish different regions and the associated isotope sequence in the graphene crystal and extract the superlattice period for each region. Regions consist of between 100 and 2000 gas pulses and result in average superlattice periods ranging from 6 to 225 nm. Figure \ref{gasflows} shows a typical gas flow sequence along with the associated isotope distribution as a function of radial distance. Growth conditions were chosen to produce isolated graphene single crystals and graphene was subsequently deposited onto Si/SiO$_2$ wafers by PMMA wet transfer for Raman spectroscopy.

\section{Characterization of graphene Isotope Superlattice}
\label{sec:char}

In order to demonstrate the successful synthesis of an isotopic superlattice we perform a careful analysis of the measured gas flow rates during CVD growth and the Raman spectra of the resulting samples. Samples were characterized by Raman mapping using a Renishaw Invia system and a 514 nm laser excitation source. 

Using the growth of a single crystal with regions of varying isotope concentration in ten percent concentration steps \cite{whit17}, we can extract the Raman G peak position and width dependence on the \ctw isotope concentration $\rho$ for homogeneous isotope mixtures. We can fit the peak with a Lorentzian of width $\gamma_\rho$ and position $\omega_\rho$, where

\begin{equation}
\omega_\rho=\omega^{12}\sqrt{\frac{12+\rho}{13}} \mbox{ and }\gamma_\rho=\gamma_0+\gamma_1\frac{f(\rho)}{f(0.5)}.
\label{exp_peak_pos}
\end{equation}

$f(\rho)$ was calculated by Rodriguez-Nieva {\it et al.} \cite{rodr12}, who found
\begin{equation}
f(\rho)=\rho(1-\rho)\,\delta m^2\,(1+\rho\delta m)^{-5/2}.
\end{equation}

$\delta m=\frac{1}{12}$ is the relative mass difference of \cth and \ctw. We find $\gamma_1= 6.2$ \cm and $\gamma_0=12.4$ \cm, in line with previously reported values\cite{rodr12,carv15} (details and figure shown in supporting information). 

We can now try to predict the expected peak structure for an inhomogeneous distribution of isotopes, such as in an isotope superlattice, where the concentration of each dominant isotope region is not necessarily 100\% pure as expected from the growth log shown in figure \ref{gasflows}. We expect the inhomogeneous peak structure to be given by a sum of Lorentzian peaks weighted by the isotope concentration distribution $P_\rho$:

\begin{equation}
I(\omega)=\sum_\rho P_\rho \frac{ \gamma_\rho/2}{(\omega-\omega_\rho)^2 +(\gamma_\rho/2)^2},
\label{exp_peak}
\end{equation}

\begin{figure*}[htbp]
\includegraphics[width=.49\columnwidth]{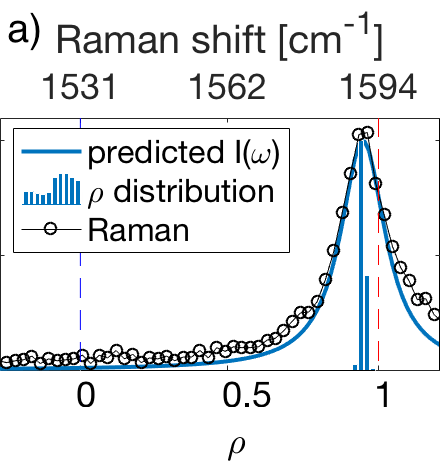}
\includegraphics[width=.49\columnwidth]{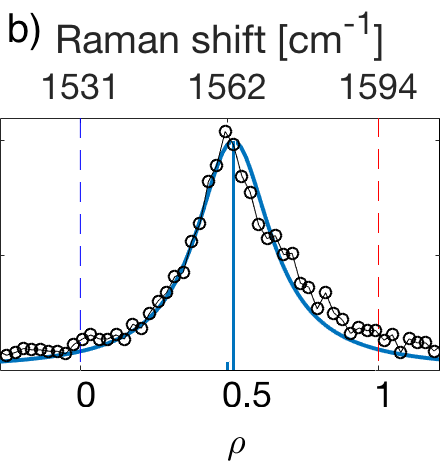}
\includegraphics[width=.49\columnwidth]{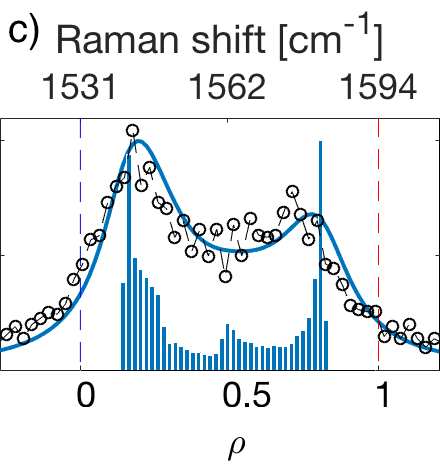}
\includegraphics[width=.49\columnwidth]{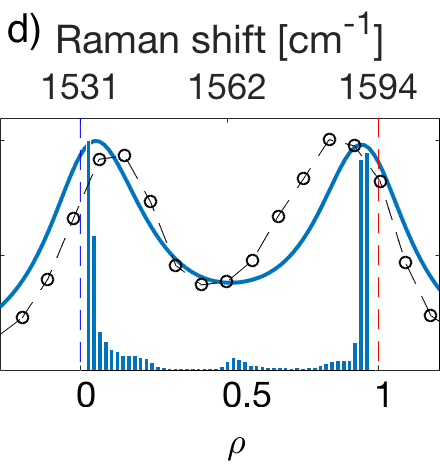}
\caption{
Predicted vs. measured G peak Raman spectra for different isotope distributions. Bar graph indicates the isotope concentration distribution extracted from the growth logs. Solid line is the predicted Raman peak shape from equation \eqref{exp_peak} and circles show the measured Raman spectra for the corresponding region. a)100\% \ctw graphene b) 50\% \ctw mix c) low isotope concentration superlattice ($C\simeq 0.7$) d) high isotope concentration superlattice ($C\simeq 0.9$).  The measured Raman spectra represent a single map point, with short collection time, and as a result are relatively noisy. These are shown as is, in order to avoid introducing extra linewidth broadening by averaging over multiple data points.}
\label{iso_distrib}
\end{figure*}

In figure \ref{iso_distrib} we compare the measured Raman G peak with the expected peak structure from equation \eqref{exp_peak} using the measured isotope distribution from the gas flows of the growth (see fig. \ref{gasflows}b). As we can see in fig. \ref{gasflows}b the peak concentrations are not exactly 100\% or 0\% in each isotope region and they depend on the growth as shown in figure \ref{iso_distrib}, which justifies the use of equation \eqref{exp_peak}. Examples of different binary distributions of concentrations are shown in figure \ref{iso_distrib} and the corresponding predicted Raman peak structure. The excellent agreement between predicted and measured spectrum indicates that the distribution of isotopes within the samples is well represented by the measured gas flows and that the Raman G peak is a determined by the corresponding isotope distribution. As such we can use the G peak position and lineshape as a measure of the superlattice purity.

For a bimodal isotope concentration distribution such as those shown in figure \ref{iso_distrib} a and b we find that we can individually resolve the \ctw and \cth Raman G peaks. We consider the simplifying approximation that the concentration distribution is well described by considering two peaks centered at $\rho_1$ and $\rho_2$ with integrated peak counts $N_1$ and $N_2$ and therefore the Raman intensity is well described by considering two peaks with Raman shift of $\omega_1$ and $\omega_2$ and integrated counts $N_1$ and $N_2$ where $\omega_{1,2}$ and $\rho_{1,2}$ are related through equation \eqref{exp_peak_pos}.

Two useful quantities describing the superlattice quality, the average carbon mass $\overline{M}$ and the average isotopic concentration $C$ can be calculated from the G peak lineshape as:

\begin{equation}
\overline{M}=13-\frac{N_1\rho_1+N_2\rho_2}{N_1+N_2}
\label{eqm}
\end{equation}

\begin{equation}
C=\frac{\omega_2-\omega_1}{2(\omega_G^{12}-\omega_G^{13})}+\frac{1}{2},
\label{eqc}
\end{equation}

\begin{figure}[htb]
\includegraphics[width=3.2in]{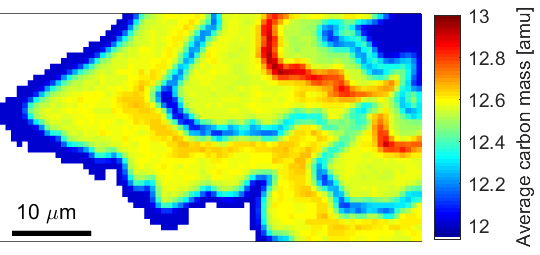}
\includegraphics[width=3.2in]{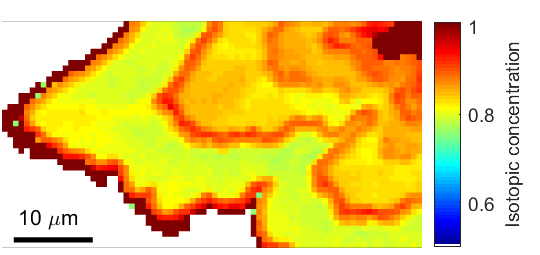}
\includegraphics[width=3.2in]{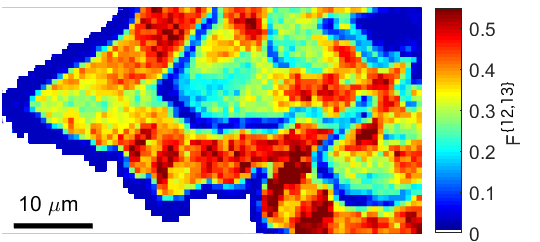}
\caption{ Raman maps of a) average carbon mass, $\overline{M}$ (equation \ref{eqm}) , b) Isotopic concentration, C (equation \ref{eqc}) calculated from the G peak position and intensity. c) $F^{\{12,13\}}$}
\label{avmass}
\end{figure}

Where $\omega_G^{12}-\omega_G^{13}\simeq 62$ \cm. In figure \ref{avmass} these two quantities are shown for a given superlattice Raman map, extracted by fitting the G peak to a double Lorentzian function. 

From the Raman maps shown in figure \ref{avmass} we observe that the sample contains several regions with a periodic superlattice structure and average mass $\sim$12.5 amu separated by lines of pure \ctw and \cth graphene. We note that the superlattice period is generally smaller than the spot size of the Raman excitation laser and as such each data point is averaging over several periods.
We are able to realize isotopic concentrations from 0.8 to greater than 0.9. This assumes constant $\omega_\rho$ which is in agreement with our numeric results for superlattice periods > 10 nm.

In general we also observe broadening of the Raman peaks compared to pristine \ctw or \cth graphene, which we attribute to increased phonon scattering from isotope impurities \cite{rodr12} and a further broadening caused by the inhomogeneous isotope distribution within a given band.

From the Raman map for each region (delimited by regions of pure \ctw or \cth graphene) the average superlattice period, $L_s$ can be calculated by measuring the length of the region and the number of isotopic methane pulses employed in the growth phase.

\section{Conclusions}

We presented the first experimentally realized Raman spectroscopy of nm scale graphene isotope superlattices. Characterization of these superlattices shows evidence of high isotopic concentration > 0.9 and small superlattice period $\sim$ 6 nm. We found a new mixed Raman process involving spatially separated phonons from both the \ctw and \cth bands. The mixed Raman process is well explained quantitatively by the real space Raman picture, involving two phonon resonant Raman processes. The intensity of this process increases as a function of the superlattice interface density and depends in the mean free path of the photo-excited carriers involved in the Raman process. We show the dependence of the photo-excited electron mean free path on substrate by comparing suspended and SiO$_2$ supported graphene where the mean free paths was found to be 7.4 nm and 18 nm respectively.

\section{Acknowledgments}
This work was supported by NSERC, FRQNT and INTRIQ.

\bibliography{phonon}
\end{document}